\documentclass[12pt,epsfig]{article}
\begin{document}
\baselineskip = 16 pt

\noindent{\large\bf Symbols and synergy in a neural code}

\bigskip\bigskip

\noindent Naama Brenner,$^1$ Steven P.
Strong,$^{1,2}$ Roland Koberle,$^{1,3}$ William
Bialek,$^1$\\
and Rob R. de Ruyter van
Steveninck$^1$
\bigskip

\noindent$^1$NEC Research Institute,
4 Independence Way,
Princeton, New Jersey 08540\\
$^2$Institute for Advanced Study,
Olden Lane,
Princeton, New Jersey 08544\\
$^3$Instituto di F\'isica de S\~ao Carlos, Caixa Postal 369\\
Universidade de S\~ao Paulo,
13560--970 S\~ao Carlos, SP Brasil

\bigskip\bigskip

\bigskip\bigskip
\bigskip\bigskip
\bigskip\bigskip\hrule
\bigskip\bigskip

Understanding a neural code requires
knowledge both of the elementary symbols that
transmit information and of the algorithm for
translating these symbols into sensory signals or 
motor actions. We show
that these questions can be separated:  the information
carried by any candidate symbol in the code---a pattern
of spikes across time or across  a population of
cells---can be measured, independent of assumptions about
what these patterns might represent.  By comparing the
information carried by a compound pattern with the
information carried independently by its parts,
we measure directly the synergy among these
parts. We illustrate the use of these  methods by
applying them to experiments on the  motion
sensitive neuron H1 of the fly's visual system,
where we confirm that two spikes close together
in time carry far more  than twice the
information carried by a single spike.   We 
analyze the sources of this synergy, and provide
evidence that pairs of spikes close together in
time  may be special symbols in the code of H1.

\vfill\newpage

\section{Introduction}

Throughout the nervous
system, information is encoded in  sequences of identical
action potentials or spikes. The representation of sense
data by these spike trains has been studied for seventy
years
\cite{adrian}, but there remain many open questions about
the structure of this code.  
A full understanding of the code requires that  we
identify its elementary symbols and that we characterize
the messages which these symbols represent.
Many different possible elementary symbols have been
considered, implicitly or explicitly, in previous work on
neural coding.  These might
be the numbers of spikes in time windows of fixed size,
or alternatively the individual spikes themselves might
be the building blocks of the code.  In cells that
produce bursts of action potentials, these bursts might be
special symbols that convey information 
in addition to that carried by single spikes.  Yet
another possibility is that patterns of spikes---across
time in one cell or across a population of cells---can
have a special significance; this last possibility has
received renewed attention as techniques emerge for
recording the activity of many neurons simultaneously.

In many methods of analysis, questions about the symbolic
structure of the code are mixed with questions about what
the symbols represent.  Thus, in trying to characterize
the feature selectivity of neurons, one often makes the
a priori choice to measure the neural response as the spike count
or rate in a fixed window.  Conversely, in trying to
assess the significance of synchronous spikes from pairs
of neurons, or bursts of spikes from a single neuron, one
might search for a correlation between these events and
some particular stimulus features. In each case
conclusions about one aspect of the code are limited by
assumptions about another aspect.  Here we show
that questions about the symbolic structure of the neural
code can be separated out and answered in an information
theoretic framework, using data from suitably designed 
experiments.  This framework allows us to address directly
the significance of spike patterns or other compound 
spiking events:  How much information is carried by 
a compound event?
Is there redundancy or synergy among the individual spikes? Are
particular patterns of spikes especially informative?  

Methods to assess the significance of spike patterns 
in the neural code share a common intuitive basis:
\begin{itemize}
\item Patterns of spikes can play a 
role in representing stimuli if and only if the
occurrence of patterns is linked to stimulus variations.
\item The patterns have a {\em special} role only if this
correlation between sensory signals and patterns is not
decomposable into separate correlations between the signals
and the pieces of the pattern, e.g. the individual spikes.
\end{itemize}
We believe that these statements are not controversial. 
Difficulties arise when we try to quantify this intuitive
picture:  What is the correct measure of
correlation?  How much correlation is
significant?  Can we make statements independent of models
and elaborate null hypotheses?

The central claim of this paper is that many of these
difficulties can be resolved by careful use of ideas from
information theory.  Shannon \cite{shannon} proved that entropy
and information provide the only measures of variability
and correlation that are consistent with simple and
plausible requirements.  Further, while it may be unclear
how to interpret, for example, a 20\% increase in
correlation between spike trains, an extra bit of
information carried by patterns of spikes means precisely
that these patterns provide a factor of two increase in
the ability of the system to distinguish among different
sensory inputs.  In this work we show see that there is a
direct method of measuring the information (in bits) carried by
particular patterns of spikes, independent of models for
the stimulus features that these patterns might
represent.  In particular, we can compare the information
conveyed by spikes patterns with the information conveyed
by the individual spikes that make up the pattern, and
determine quantitatively whether the whole is more or less
than the sum of its parts.

While this method allows us to compute unambiguously how
much information is conveyed by particular patterns, it
does not tell us what this information is.  Making
the distinction between two
questions, the structure of the code and the algorithm 
for translation, we only answer the first of these two.
We believe that finding the structural properties of a
neural code independent of the translation algorithm is an essential
first step towards understanding anything beyond the single spike 
approximation. The need for identifying the elementary symbols
is especially clear when complex
multi neuron codes are considered. Moreover, a quantitative measure of the 
information carried by compound symbols will be useful in the next
stage of modeling the encoding algorithm, as a control for the validity
of models.

\section{Formalism}

In the framework of information theory \cite{shannon}
signals are generated by a source with a fixed probability
distribution, and  encoded into messages by a channel. The
coding is probabilistic, and the joint distribution of
signals and coded messages determines all quantities of
interest; in particular the information transmitted by the
channel about the source is an average over this joint
distribution. In studying a sensory system, the signals
generated by the source are the stimuli presented to the
animal,  and the messages in the communication channel are
sequences of spikes in a neuron or in a
population of neurons.  Both the stimuli and the
spike trains are random variables, and they
convey information mutually because they are
correlated. The problem of quantifying this
information has been discussed from several
points of view \cite{MacKay52,Optican87,Spikes97,Strong98}. 
Here we address the question of how much
information is carried by particular ``events'' or combinations
of action potentials.

\subsection{Defining information from one event}

A discrete event
$E$ in the spike train is defined as a  specific
combination of spikes. Examples are a single spike, 
a pair of spikes separated by a given 
time, spikes from two neurons that occur in synchrony, 
and so on.  Information is carried by the
occurrence of  events  at particular times
and not others, implying that they are correlated
with some stimulus features and not with
others.  Our task is to express this information in terms
of quantities that are easily measured experimentally.

In experiments, as in nature, the animal
is exposed to stimuli at each instant of
time.  We can describe this sensory input
by a function $s(t')$, which may have many
components to parameterize the time
dependence of multiple stimulus features.
In general, the information gained about $s(t')$ by
observing a set of neural responses is
\begin{eqnarray}
I
&=&
\sum_{\rm responses}\int Ds(t') P[s(t') \,\&\, {\rm response}]
\log_2\left(
{{P[s(t') \,\&\, {\rm response}]}
\over{P[s(t')]P[{\rm response}]}}
\right),
\nonumber\\&&
\end{eqnarray}
where information is measured in bits.
It is useful to note that this mutual information can be
rewritten in two complementary forms:
\begin{eqnarray}
I
&=&
-\int Ds(t') P[s(t')]\log_2P[s(t')]
\nonumber\\
&&\,\,\,\,\,+
\sum_{\rm responses}
P[{\rm response}]
\int Ds(t') P[s(t')|{\rm response}]\log_2P[s(t'
)|{\rm response}]\nonumber\\
&=&
S[P(s)]  -  \langle S[P(s|{\rm response})] \rangle_{\rm
response},
\label{stimspace}
\end{eqnarray}
or
\begin{eqnarray}
I &=&
-\sum_{\rm responses} P({\rm response})\log_2 P({\rm
response})
\nonumber\\
&&\,\,\,\,\,+
\int Ds(t') P[s(t')]
\sum_{\rm responses} P[{\rm response}| s(t')]\log_2
P[{\rm response}|s(t')]\nonumber\\
&=&
S[P({\rm response})] - \langle S[P({\rm response}|s)]\rangle_s,
\label{spikespace}
\end{eqnarray}
where $S$ denotes the entropy of a distribution; 
by $\langle \cdots\rangle_s$ we mean
an average over all possible values of the sensory stimulus,
weighted by their
probability of occurrence, and similarly for $\langle
\cdots\rangle_{\rm responses}$. In the first form, Eq.
(\ref{stimspace}),  we focus on what the responses are
telling us about the sensory stimulus 
\cite{Ruyter88}: different responses
point more or less reliably to different signals, and our
average uncertainty about the sensory signal is reduced by
observing the neural response.  In the second form, Eq.
(\ref{spikespace}),  we focus on the variability and
reproducibility of the neural response
\cite{Ruyter97,Strong98}. The range of
possible responses provides the system with a capacity to
transmit information, and the variability which remains
when the sensory stimulus is specified constitutes noise;
the difference between the capacity and the noise is the
information.

We would like to apply this second form to 
the case where the neural response is a particular type 
of event.  When we observe an
event $E$, information is carried by the fact that it
occurs at some particular time $t_E$.  The
range of possible responses is then the range of times
$0 < t_E < T$ in our observation window.
Alternatively, when we observe the response in a particular
small time bin of size $\Delta t$, information is carried
by the fact that the event $E$ either occurs or does not.
The range of possible responses then includes just two
possibilities.  Both of these  points of view have an
arbitrary element:  the choice of bin size
$\Delta t$ and the window size $T$.  Characterizing the
properties of the system, as opposed to our observation power,
requires taking the limit of high time resolution ($\Delta t
\rightarrow 0$) and long observation times ($T\rightarrow
\infty$). As will be shown in the next section, 
in this limit the two points of view give the
same answer for the information carried by an 
event. 

\subsection{Information and event rates}

A crucial role is played by the {\em event rate}
$r_E(t)$, the probability per unit time
that an event of type
$E$ occurs at time $t$, given the stimulus
history $s(t')$.   Empirical
construction of the event rate $r_E(t)$ requires repetition of
same stimulus history many times, so that a histogram
can be formed (see Figure 1).
For the case where events are single
spikes, this is the familiar time dependent
firing rate or post-stimulus time histogram
(Figure 1c); the generalization to other types of
events is illustrated by Figs. 1c and 1d.
Intuitively, a uniform event rate
implies that no information  is transmitted, whereas the
presence of sharply defined features in the event rate
implies that much information is transmitted by
these events; see, for example, the discussion
by Vaadia et al. \cite{Vaadia95}.
We now formalize this intuition, and show how the
average information carried by a single event is related
quantitatively to the time dependent event rate.

Let us take the first point of view about the neural response 
variable, in which the range of
responses is described by the possible arrival times $t$
of the event. What is the probability of finding the event
at a particular time $t$?
Before we know the 
stimulus history
$s(t')$, all we can say is that the
event can occur anywhere in our experimental window of
size $T$, so that the probability is uniform
$P({\rm response})\!=\!P(t)\!=\!1/T$, with an entropy of 
$S[P(t)]\!=\!\log_2 T$. Once we know the stimulus, we also know
the event rate $r_E(t)$, and so our uncertainty
about the occurrence time of the event is reduced. 
Events will occur
preferentially at times where the event rate is large, so
the probability distribution should be proportional to
$r_E(t)$; with proper  normalization 
$P({\rm response}|s)\!=\!P(t|s)\!=\!r_E(t)/(T\bar r_E)$.  Then the
conditional entropy is
\begin{eqnarray}
S[P(t|s)] &=& -\int_0^T dt~ P(t|s) \log_2 P(t|s) \nonumber \\
&=& - {1\over T}\int_0^T dt~{{r_E(t)}\over{\bar r_E}}
\log_2 \left({{r_E(t)}\over{{\bar r_E}T}}\right) .
\end{eqnarray}
In principle one should average this quantity over different
stimuli $s(t')$, however if the time $T$ is long enough and the
stimulus history sufficiently rich, it is self-averaging.
The reduction in entropy is then the gain in information, so
\begin{eqnarray}
I(E;s) &=& 
S[P(t)]-S[P(t|s)]
\nonumber \\
&=& 
{1\over T}
\int_0^T dt\, \left({{r_E(t)}
\over{\bar r_E}}\right)
\log_2
\left(
{{r_E(t)}
\over{\bar r_E}}
\right) . \label{final1eventinfo}
\end{eqnarray}
This formula expresses the information conveyed by an event of type $E$
as an integral over time, of a quantity which depends only
on the responses. There is no explicit dependence on the joint
distribution of stimuli and responses; it is implicit that by
integrating over time we are in fact sampling the distribution of
stimuli, whereas by estimating the function $r_E(t)$ as a histogram
we are sampling the
distribution of the responses given a stimulus. 
We may write the
information equivalently as an average over the stimulus instead of over time,
\begin{eqnarray}
I(E;s) &=& \Bigg\langle \left(
{{r_E(t)}
\over{\bar r_E}} \right)
\log_2 \left({{r_E(t)}\over{\bar r_E}}
\right) \Bigg\rangle_s,  \label{1eventinfo}
\end{eqnarray}
where here the average is over all possible value of $s$ weighted by 
their probabilities $P(s)$.

In the second view, the neural response
is a  binary random variable,
$\sigma_E \in \{0,1\}$, marking the occurrence or non occurrence of
an event of type $E$ in a small time bin of size $\Delta
t$. Suppose, for simplicity, that 
the stimulus takes on a finite set of values $s$ with
probabilities
$P(s)$.
These in turn induce the event $E$ 
with probability $p_E(s)\!=\!P(\sigma_E\!\!=\!\!1|s)\!=\!r_E(s)\Delta t$, with
an average probability for the occurrence of the event $\bar{p}_E
=\sum_s P(s) ~r_E(s) \Delta t =
{\bar r}_E  \Delta t$. The information is the
difference between the prior entropy and the
conditional entropy: 
$I(E;s)\!=\!S(s)\!-\!\langle S(s|\sigma_E ) \rangle$, 
where the conditional entropy is an
average over the two possible values of $\sigma_E$.
The conditional probabilities are found from Bayes' rule, 
\begin{eqnarray}
{\mbox
P}(s|\sigma_E\!=\!1)&=&\frac{P(s) p_E(s)}{\bar{p}_E}\nonumber \\
{\mbox
P}(s|\sigma_E\!=\!0)&=&\frac{P(s)(1-p_E(s))}{(1-\bar{p}_E)},
\end{eqnarray}
and with these one finds the information,
\begin{eqnarray}
I(E;s) &=&  -\sum_s P(s)\log_2 P(s) + 
\sum_{\sigma_E=0,1} P(s|\sigma_E) \log_s P(s|\sigma_E)
\nonumber \\
\!\!\!\!&=&\sum_s P(s) \left[
p_E(s)~\log_2\Bigg(\frac{p_E(s)}{\bar{p}_E}\Bigg) +
(1\!-\!p_E(s))\log_2\Bigg(\frac{1\!-\!p_E(s)}
{1\!-\!\bar{p_E}} \Bigg) \right] .
\label{discr_1info}
\end{eqnarray}
This expression is again an average over all stimulus values,
of a property which only depends on the responses. 
Taking the limit $\Delta t\to 0$, consistent with the requirement that
the event can occur at most once, one finds the average
information conveyed 
in a small time bin; dividing by the average 
probability of an event
one obtains 
Eq. (\ref{1eventinfo}) as the information
per event.  

Equation (\ref{1eventinfo}), and its time averaged form
(\ref{final1eventinfo}), is an exact formula  
which can be used in any situation where a  rich stimulus
history can be presented repeatedly.
It enables the evaluation of the information  for
arbitrarily complex events, independent of assumptions
about the encoding algorithm. 

Let us consider in more detail the 
simple case where events are single
spikes. Then the average information
conveyed by a single spike becomes an
integral over the time dependent spike rate
$r(t)$,
\begin{equation}
I({\rm 1\,\, spike;s})
=
{1\over T}
\int_0^T dt\, \left({{r(t)}
\over{\bar r}}\right)
\log_2
\left(
{{r(t)}
\over{\bar r}}
\right) .
\label{final1spikeinfo}
\end{equation}
It makes sense that the
information carried by single spikes should be related to the
spike rate, since this rate as a function of time gives a complete
description of the `one body' statistics of
the spike train, in the same way that the single particle
density describes the one body statistics of a gas or
liquid. Several previous works have noted this
relation, and the
formula (\ref{final1spikeinfo}) 
has an interesting history. If the spike
train is a modulated Poisson process, then  Eq.
(\ref{final1spikeinfo}) provides an upper bound on
information transmission (per spike) by the spike train as
a whole \cite{Bialek90}.  In studying the coding of location
by cells in the rat hippocampus, Skaggs et al. \cite{Skaggs93}
assumed that successive spikes carried independent
information, and that the spike rate was determined by
the instantaneous location, and obtained   Eq.
(\ref{final1spikeinfo}) with the time average replaced by
an average over locations. DeWeese \cite{DeWeese96} showed that
the rate of information transmission by a spike train
could be expanded in a series of integrals over
correlation functions, where successive terms would be
small if the number of spikes per correlation time were
small; the leading term, which would be exact if spikes
were uncorrelated, is Eq. (\ref{final1spikeinfo}). 
Panzeri et al. \cite{Panzeri96} show that Eq.
(\ref{final1spikeinfo}), multiplied by the mean spike rate
to give an information rate (bits/s), is the correct
information rate if we count spikes in very brief segments
of the neural response, which is equivalent to asking for
the information carried by single spikes.  For further 
discussion of the relation to previous work, see Appendix A.

A crucial point
here is the generalization to Eq.
(\ref{final1eventinfo}), and this result applies to the
information content of {\em any} point events in the
neural response---pairs of spikes with a certain
separation, coincident spikes from two cells, ... ---not
just single spikes.  
Moreover, in the analysis of
experiments we will emphasize the use of this formula
as an exact result for the information content of
single events, rather than an approximate result for
the spike train as a whole, and this approach will enable
us to address questions concerning the structure of the code
and the role played by various point events.

\section{Experiments in the fly visual system}

In this section, we use our formalism to analyze 
experiments on the movement sensitive cell H1 in the visual
system of the blowfly {\em Calliphora vicina}. We address the issue of
the information conveyed by pairs of spikes in this neuron, as
compared to the information conveyed independently by single spikes.
The quantitative results of this section ---numbers for the information,
effects of synergy and redundancy among spikes--- are specific to
this system and to the stimulus conditions used. The theoretical
approach, however, is valid generally and can be applied 
similarly to other experimental systems, to find out the significance of
various patterns in single cells or in a population.

\subsection{Synergy between spikes}

The experimental setup described in Appendix B gives us control over
the input and output of the H1 neuron in the fly.
The horizontal motion across the visual field
is the input sensory stimulus $s(t)$, which
we draw from a probability distribution $P(s)$, and the spike train
recorded from H1 is the neural response.
Figure 1a shows a segment of the stimulus 
presented to the fly, and 1b illustrates 
the response to many repeated presentations of this segment.
The histogram of spike times across the ensemble of repetitions provides
an estimate of the spike rate  $r(t)$ (Fig. 1c), and  
Eq. (\ref{final1eventinfo})
gives the information  carried by a single spike, $I({\rm 1\ spike};s) 
= 1.53 \pm
0.05\ {\rm bits}$.  Figure 2 illustrates the details of how the formula
was used, with an emphasis on the effects of finiteness of the data. 
In this experiment, a stimulus of length $T=10$ sec was repeated 350
times. As seen from Figure 2, a stable result could be obtained
from a smaller number of repetitions.

If each spike were to convey information independently, then with the
mean spike rate $\bar r = 37$ spikes/s, the total information rate would
be $R_{\rm info} = 56$ bits/s.  We used the variability and
reproducibility of continuous segments in the neural response 
\cite{Strong98}, in order to estimate the total information rate in the
spike train in this experiment, and found that
$R_{\rm info} = 75$ bits/s.   
Thus, the information conveyed by the spike
train as a whole is larger than the sum of 
contributions from individual spikes, indicating cooperative
information transmission by patterns of spikes
in time.
This synergy among spikes motivates the search for
especially informative patterns in the spike train.

We consider compound events that consist of
two spikes separated by a time $\tau$, 
with no constraints on what happens between
them.  
Figure 1 shows segments of the event rate $r_\tau (t)$ for $\tau = 3 
(\pm 1)$ ms (Fig. 1d), and for $\tau = 17 (\pm 1)$ ms (Fig. 1e). 
The information carried by spike pairs as a function of the 
interspike time $\tau$, computed from Eq. 
(\ref{final1eventinfo}), is shown in Fig. 3.  For large $\tau$ 
spikes contribute 
independent information, as expected.  This independence is 
established within $\sim 30 - 40$ ms, comparable to the 
behavioral response times of the fly \cite{Land74}. There 
is a mild redundancy ($\sim 10 -  20\%$) at intermediate 
separations, and a very large synergy
(up to $\sim\!\! 130\%$) at small $\tau$.

Related results were obtained using
the correlation of spike patterns with 
stimulus features \cite{Ruyter88}.
There the information carried by spike
patterns was estimated from the distribution of stimuli
given each pattern, thus constructing a statistical model
of what the patterns ``stand for'' (see details
in Appendix A).   
Since the time dependent stimulus is in general of high
dimensionality, its distribution cannot be sampled directly
and some approximations must be made.  de Ruyter 
van Steveninck and Bialek \cite{Ruyter88} made the 
approximation that patterns of a few
spikes encode projections of the stimulus onto low
dimensional subspaces, and the information carried by such
patterns was evaluated only in this subspace. 
The informations obtained in this approximation
are bounded from above by the true
information carried by the patterns, as estimated directly
with the methods presented here.  

\subsection{Origins of synergy}

Synergy means, quite literally, 
that two spikes together tell
us more than two spikes separately. 
Synergistic coding is often discussed for populations of cells,
where extra information is conveyed by patterns of
coincident spikes from several neurons
\cite{Abeles93,Hopfield95,Meister95a,Singer95}, 
while here we see
direct evidence for extra information in pairs
of spikes across time.  The mathematical framework for describing these
effects is the same, and a natural question is:
what are the conditions for synergistic coding?

The average synergy ${\rm Syn}[E_1, E_2;s]$  between two events
$E_1$ and $E_2$ is
the difference between the information about the stimulus $s$
conveyed by the pair, and the information conveyed by 
the two events independently,
\begin{equation}
{\rm Syn}[E_1, E_2; s] = I[E_1 , E_2  ;s] - 
(I[E_1 ; s] + I[E_2 ;s]).
\label{redund-def}
\end{equation}
We can rewrite the synergy as:
\begin{equation}
{\rm Syn}[E_1, E_2 ; s] = 
I[E_1;E_2| s] - I[E_1;E_2].
\label{redund-result}
\end{equation}
The first term is the mutual  information between
the  events computed across an ensemble  of
repeated presentations of the same stimulus
history. It describes the
gain in information due to the locking of 
compound event $(E_1,E_2)$ to
particular stimulus features.   If events $E_1$
and
$E_2$  are correlated individually with the
stimulus but not with one another, this term will
be zero, and these events cannot be synergistic
on average. The second term is the mutual
information between events when  the stimulus is
not constrained, or equivalently the
predictability of event $E_2$ from $E_1$.  This
predictability limits the capacity of $E_2$ to
carry information beyond that already conveyed by
$E_1$. Synergistic coding (${\rm Syn} > 0$) 
thus requires that the mutual information among the spikes is 
increased by specifying the stimulus, which makes precise the 
intuitive idea of `stimulus dependent correlations'.

Returning to our experimental example, 
we identify the 
events $E_1$ and $E_2$ as the arrivals of two spikes, 
and consider the synergy between  them as a function of
the time  $\tau$  between them. In terms of event rates,
we compute the information carried by a pair of spikes 
separated by a time $\tau$, 
Eq. (\ref{final1eventinfo}), as
well as the information carried by two individual 
spikes. The difference between these two quantities  is the
synergy between  two spikes, which can be written as 
\begin{eqnarray} 
{\rm Syn}(\tau) &=& - \log_2
\left(
{ {{\bar r}_\tau} \over {{\bar r}^2} }
\right) 
+ 
{1\over T}
\int_0^T dt 
{ {r_\tau (t)} \over {{\bar r}_\tau} }
\log_2 
\left[
{ {r_\tau (t)} \over {r (t)r(t-\tau)}}\right]\nonumber\\
&&\,\,\,\,\,
+ {1\over T}\int_0^T dt 
\left[ 
{ {r_\tau (t)} \over {{\bar r}_\tau} }  
 + 
{ {r_\tau (t+\tau)}\over {{\bar r}_\tau} } - 
 2{ {r(t)} \over {\bar r} }
\right] \log_2[r(t)] .
\end{eqnarray}
The first term in this equation is the logarithm of the
normalized correlation  function, and hence measures
the rarity of spike pairs with  separation $\tau$; the
average of this term over $\tau$ is the mutual information
between events in Eq. (\ref{redund-result}). The second
term is related to the local correlation function and 
measures the extent to which the stimulus modulates the 
likelihood of spike pairs. The average of this term over
$\tau$  gives the mutual information conditional on
knowledge of the  stimulus [the first term in Eq.
(\ref{redund-result})]. The  average of the third term
over $\tau$ is zero, and numerical  evaluation of this
term from the data shows that it is  negligible at most
values of $\tau$.

We thus find that the synergy between spikes is approximately a sum of
two terms, whose averages over $\tau$ are the terms in Eq. 
(\ref{redund-result}).
A spike pair with a separation $\tau$
then has two types of contributions to
the  extra information it carries: 
the two spikes can be correlated conditional on
the stimulus, or the pair could be a rare and thus
surprising event.
The rarity of brief pairs is related to neural
refractoriness, but
this effect alone is insufficient to enhance information
transmission; the rare events must also be related reliably to the stimulus.
In fact, conditional on the stimulus, the  spikes
in rare pairs  are strongly correlated with each
other, and this is visible in Fig. 1a:  
from trial to trial, adjacent spikes jitter together
as if connected by a stiff spring.  To quantify this effect, we find for each
spike in one trial the closest spike in successive trials, and measure
the variance of the arrival times of these spikes.  Similarly,
we measure the variance of the interspike times. Figure 4a  shows the
ratio of the interspike time variance 
to the sum of the arrival time variances of the spikes
that make up the pair. For large separations this ratio is unity,
as expected if spikes are locked independently to the stimulus, but as
the two spikes come closer it falls below one quarter.

Both the conditional correlation among the members of the pair (Fig. 4a)
and the relative synergy (Fig. 4b) 
depend strongly on the interspike
separation.  This dependence is nearly invariant to changes in 
image contrast, although the spike rate and other statistical 
properties are strongly affected by such changes. 
Brief spike pairs seem to
retain their identity as specially informative symbols over a
range of input ensembles.
If particular
temporal patterns are especially informative, then
we would lose information if we
failed to distinguish among different patterns. 
Thus there are two notions of time resolution for spike pairs:  the
time resolution with which the interspike time is defined,
and the absolute time resolution
with which the event is marked.  
Figure 5 shows that, for small interspike times,
the information is much
more sensitive to changes in the interspike time resolution (open
symbols) than
to the absolute time resolution (filled symbols). 
This is related to
the slope in Figure 2: in regions where the slope is large, events should
be finely distinguished in order to retain the information.

\subsection{Implications of synergy}
The importance of spike timing in the neural
code has been under debate for some time now. 
We believe that some issues in this debate
can be clarified using a direct information theoretic
approach. 
Following MacKay and McCulloch \cite{MacKay52},
we know that marking spike arrival times with
higher resolution provides an increased capacity
for information transmission.
The work of Strong et al. \cite{Strong98} shows that 
for the fly's H1 neuron, the
increased capacity associated with  
spike timing is indeed used with nearly constant efficiency
down to millisecond resolution.  
This efficiency can be the result of a tight
locking of individual spikes to a rapidly varying stimulus,
and it could also be the result of 
temporal patterns providing information beyond
rapid rate modulations.  The analysis given here shows
that for H1, pairs of spikes
can provide much more information than two individual
spikes, information transmission is much more sensitive
to the relative timing of spikes than to their
absolute timing, and these synergistic effects survive 
averaging over all similar patterns in an experiment.
On the
time scales of relevance to fly behavior, the amount of synergy
among spikes in H1 allows this single cell to provide an
extra factor of two in resolving power for distinguishing
different trajectories of motion across the visual
field.

\section{Summary}

In summary, information theory allows us  to
quantify  the symbolic structure  of a neural
code  independent of the rules for
translating between spikes and stimuli.  
In particular, this approach tests directly
the idea that patterns of spikes are  special events in
the code,  carrying more
information than expected by adding the  contributions
from individual spikes.  These quantities can be measured
directly from  data.
It is of practical importance that the formulas rely on
low order statistical measures of  the neural response,
and hence do not require enormous data sets to reach 
meaningful conclusions.  The method is of general validity
and is applicable to patterns of spikes across a
population of neurons, as well as across time.

In our experiments on the fly visual system, we found that
an event composed of 
a pair of spikes can carry far more than the information
carried independently by its parts.
Two spikes that occur in rapid succession appear to
be special symbols that have an integrity
beyond the locking of individual spikes to the
stimulus. This is analogous to the encoding of
sounds in written English: the  symbols `th,'
`sh,' and `ch' are each elementary and stand for
sounds that are not decomposable into sounds
represented by each  of the constituent letters.
For such pairs to act effectively as special
symbols, mechanisms for `reading' them must exist
at subsequent levels of processing. Synaptic
transmission is sensitive to interspike times in
the 2 -- 20 ms range
\cite{Magelby87},  and it is natural to suggest that
synaptic mechanisms on this time scale play a
role in such reading. Recent work on the
mammalian visual system  \cite{Usrey98} provides
direct evidence that pairs of spikes close
together in time can be especially efficient in
driving postsynaptic neurons.

\section*{Acknowledgements}

We thank G. Lewen and A. Schweitzer for their help with the experiments and
N. Tishby for many helpful discussions.
Work at the IAS was supported in part by DOE grant DE--FG02--90ER40542,
and work at the IFSC was supported by the Brazilian agencies FAPESP and CNPq.
\newline
\newline
\newline
\setcounter{equation}{0}
\def\theequation{A.\arabic{equation}}
{\large\bf Appendix A: Relation to previous work}\\ \\
Patterns of spikes and their relation to
sensory stimuli have been quantified in the
past through the use of correlation functions.
The event rates that we have defined here, which
are directly connected to the information carried by
patterns of spikes by Eq. (\ref{final1eventinfo}), 
are in fact just properly normalized
correlation functions.  The event rate for pairs of
spikes from two separate neurons is related to the
joint post-stimulus time histogram defined by Aertsen and
coworkers \cite{Aertsen89,Vaadia95}
Making this connection explicit is also an opportunity to
see how the present formalism applies to events defined
across two cells.

Consider two cells, $A$ and $B$, generating spikes at
times $\{ t_{\rm i}^A\}$ and $\{ t_{\rm i}^B\}$,
respectively.  It will be useful to think of the spike
trains as sums of unit impulses at the spike times,
\begin{eqnarray}
\rho^A(t) &=& \sum_{\rm i} \delta (t - t_{\rm i}^A)
\\
\rho^B(t) &=& \sum_{\rm i} \delta (t - t_{\rm i}^B) .
\end{eqnarray}
Then the time dependent spike rates for the two cells
are
\begin{eqnarray}
r^A (t)  &=&
\langle \rho^A (t)\rangle_{\rm trials} ,\\
r^B(t)  &=&
\langle \rho^B (t)\rangle_{\rm trials} ,
\end{eqnarray}
where $\langle \cdots \rangle_{\rm trials}$ denotes an
average over multiple trials in which the same time
dependent stimulus $s(t')$ is presented.  
These spike rates are the probabilities per unit time for
the occurrence of a single spike in either cell $A$ {\em
or} cell $B$, also called the post-stimulus time
histogram (PSTH).  We can define the probability per unit
time for a spike in cell $A$ to occur at time $t$ {\em
and} a spike in cell $B$ to occur at time $t'$, and this
will be the joint post-stimulus time histogram,
\begin{equation}
{\rm JPSTH}^{AB} (t,t') = \langle \rho^A (t)
\rho^B (t') \rangle_{\rm trials} .
\end{equation}
Alternatively, we can consider an event $E$ defined by a spike
in cell $A$ at time $t$ and a spike in cell $B$ at time
$t-\tau$, with the relative time $\tau$ measured to a
precision of $\Delta\tau$.  Then the rate of these events is
\begin{eqnarray}
r_E(t)
&=& 
\int_{-\Delta\tau/2}^{\Delta\tau/2}
dt' ~{\rm JPSTH}^{AB} (t,t-\tau + t')\\
&\approx&
\Delta\tau {\rm JPSTH}^{AB} (t,t-\tau ),
\end{eqnarray}
where the last approximation is valid if our time
resolution is sufficiently high.   Applying our general
formula for the information carried by single events, Eq.
(\ref{final1eventinfo}), the information carried by pairs of
spikes from two cells can be written as an integral over
diagonal ``strips'' of the ${\rm JPSTH}$ matrix,
\begin{equation}
I(E;s)
=
{1\over T}
\int_0^T dt 
{{{\rm JPSTH}^{AB} (t,t-\tau )}\over
{\langle {\rm JPSTH}^{AB} (t,t-\tau )\rangle_t}}
\log_2
\left[
{{{\rm JPSTH}^{AB} (t,t-\tau )}\over
{\langle {\rm JPSTH}^{AB} (t,t-\tau )\rangle_t}}
\right] ,
\label{JPSTHinfo}
\end{equation}
where $\langle {\rm JPSTH}^{AB} (t,t-\tau )\rangle_t$
is an average of the ${\rm JPSTH}$ over time, which is
equivalent to the standard correlation function between
the two  spike trains.

The discussion by Vaadia et al. \cite{Vaadia95} emphasizes that
modulations of the $\rm JPSTH$ along the diagonal strips
allows correlated firing events to convey information
about sensory signals or behavioral states, and this
information is quantified by Eq. (\ref{JPSTHinfo}).
The information carried by the individual cells is related to the
corresponding integrals over spike rates, Eq. (\ref{final1spikeinfo}).
The difference between the the information conveyed by the 
compound spiking events $E$, and the informations conveyed by 
spikes in the two cells independently,
is precisely the synergy between the two cells
at the given time lag $\tau$. For $\tau\!=\!0$, it is the
synergy --or extra information-- conveyed by synchronous
firing of the two cells.

We would like to connect the present approach also with
previous work which focused on how events
reduce our uncertainty about the stimulus \cite{Ruyter88}.
Before we observe the
neural response, all we know is that stimuli are chosen
from a distribution $P[s(t' )]$.  When we observe an
event $E$ at time $t_E$, this should tell us something
about the stimulus in the neighborhood of this time, and
this knowledge is described by the conditional
distribution $P[s(t' )|t_E]$.   If we go back to the
definition of the mutual information between responses and
stimuli, we can write the information conveyed by one
event in terms this conditional distribution,
\begin{eqnarray}
I(E;s)
&=&
\int Ds(t') \int dt_E P[s(t') , t_E]
\log_2 \left(
{{P[s(t') , t_E]}\over{P[s(t') ] P[t_E]}}
\right) \nonumber \\ 
&=&
\int dt_E P[t_E]
\int Ds(t') P[s(t') | t_E]
\log_2
\left(
{{P[s(t')| t_E]}\over{P[s(t')]}}
\right) .
\label{rceinfo}
\end{eqnarray}
If the system is stationary then the coding
should be invariant under time translations:  
\begin{equation}
P[s(t') | t_E ] = P[s(t' + \Delta t' ) | t_E +
\Delta t'] .
\end{equation}
This invariance means that the integral over stimuli in Eq.
(\ref{rceinfo}) is independent of the event arrival time
$t_E$, so we can simplify our expression for the
information carried by a single event,
\begin{eqnarray}
I(E;s)&=&
\int dt_E P[t_E]
\int Ds(t') P[s(t') | t_E]
\log_2
\left(
{{P[s(t')| t_E]}\over{P[s(t')]}}
\right) \nonumber\\
&=&
\int Ds(t') P[s(t') | t_E]
\log_2
\left(
{{P[s(t')| t_E]}\over{P[s(t')]}}
\right) .
\label{88info}
\end{eqnarray}
This formula
was used by de Ruyter van Steveninck and Bialek \cite{Ruyter88}
To connect with the present work, we express the information
in Eq. (\ref{88info}) as an average over the stimulus,
\begin{eqnarray}
I(E;s)~=~ \Bigg\langle 
\left( {{P[s(t')| t_E]}\over{P[s(t')]}} \right) 
\log_2
\left( {{P[s(t')| t_E]}\over{P[s(t')]}} \right)
\Bigg\rangle_s.
\end{eqnarray}
Using Bayes' rule,
\begin{equation}
{{P[s(t') | t_E]}\over {P[s(t')]}} =
{{P[t_E | s(t') ]}\over{P[t_E ]}}  = 
{{r_E(t_E)}\over{{\bar r}_E}}
\end{equation}
where the last term is a result of the 
distributions of event arrival times being
proportional to the event rates,
as defined above.  Substituting the back to Eq. (\ref{88info}),
one finds the equivalent of Eq. (\ref{1eventinfo}).
\newline
\newline
\newline
\setcounter{equation}{0}
\def\theequation{B.\arabic{equation}}
{\large\bf Appendix B: Experimental setup}\\ \\
In the experiment we used a female blowfly, which was a first
generation offspring of a wild fly caught outside. The
fly was put inside a plastic tube and immobilized with
wax, with the head protruding out. The proboscis was left
free so that the fly could be fed regularly with some sugar
water. A small hole was cut in the back of the head, close to the
midline on the right side. Through this hole, a tungsten
electrode was advanced into the lobula plate. This area, which is
several layers back from the compound eye, includes a group of large
motion detector neurons with wide receptive fields and strong direction
selectivity. We recorded spikes
extracellularly from one of these, the contralateral H1 neuron 
\cite{Facets89}.   The electrode
was positioned such that spikes from H1 could be discriminated
reliably, and converted into TTL pulses by a simple threshold
discriminator. The TTL pulses fed into a CED 1401 interface,
which time stamped the digitized spikes at 10 $\mu s$ resolution.
To keep exact synchrony over the duration of the experiment, the
spike timing clock was derived from the same internal CED 1401
clock that defined the frame times of the visual stimulus.

The stimulus was an rigidly moving bar pattern, displayed on a Tektronix
608 high brightness display. 
The radiance at average intensity was about
20 mW/(m$^2$ $\cdot$ sr), which
amounts to about 5 $\cdot 10^4$ effectively transduced
photons per photoreceptor per second \cite{Dubs84}.
The bars were oriented vertically, 
with intensities chosen at random to be ${\bar I}(1\pm C)$, 
where $C$ is the contrast. 
The distance between the fly and the screen
was adjusted so that angular subtense of a bar
equaled the horizontal interommatidial angle in the stimulated
part of the compound eye. 
This setting was 
found by determining the eye's spatial Nyquist
frequency through the reverse reaction \cite{Gotz64}.
For this fly, the horizontal interommatidial
angle was 1.45$^{\circ}$, and the distance to the screen 
105 mm. 
The fly viewed the display through a round 80 mm
diameter diaphragm, showing approximately 30 bars. 
From this we estimate the number of
stimulated ommatidia in the eye's hexagonal raster to be about
612.

Frames of the stimulus pattern were refreshed every 2 ms, and 
with each new frame
the pattern was displayed at a new position.  This 
resulted in an apparent horizontal motion of the bar pattern, 
which is suitable to excite the H1 neuron.
The pattern position was defined by a pseudorandom sequence, simulating
independent random numbers uniformly distributed between
-0.47$^{\circ}$ to + 0.47$^{\circ}$ (equivalent to -0.32 to +0.32 omm,
horizontal ommatidial spacings). This corresponds to a diffusion
constant of 18.1($^{\circ}$)$^2$/s or 8.6 omm$^2$/s.
The sequence of pseudorandom numbers contained
a repeating part and a nonrepeating part, each 10 seconds long,
with the same statistical parameters. 
Thus in each 20
second cycle the fly saw a 10 second movie that it had seen 20 seconds
before, followed by a 10 second movie that was generated independently.

\newpage
\leftline{\large\bf Figures}
\bigskip

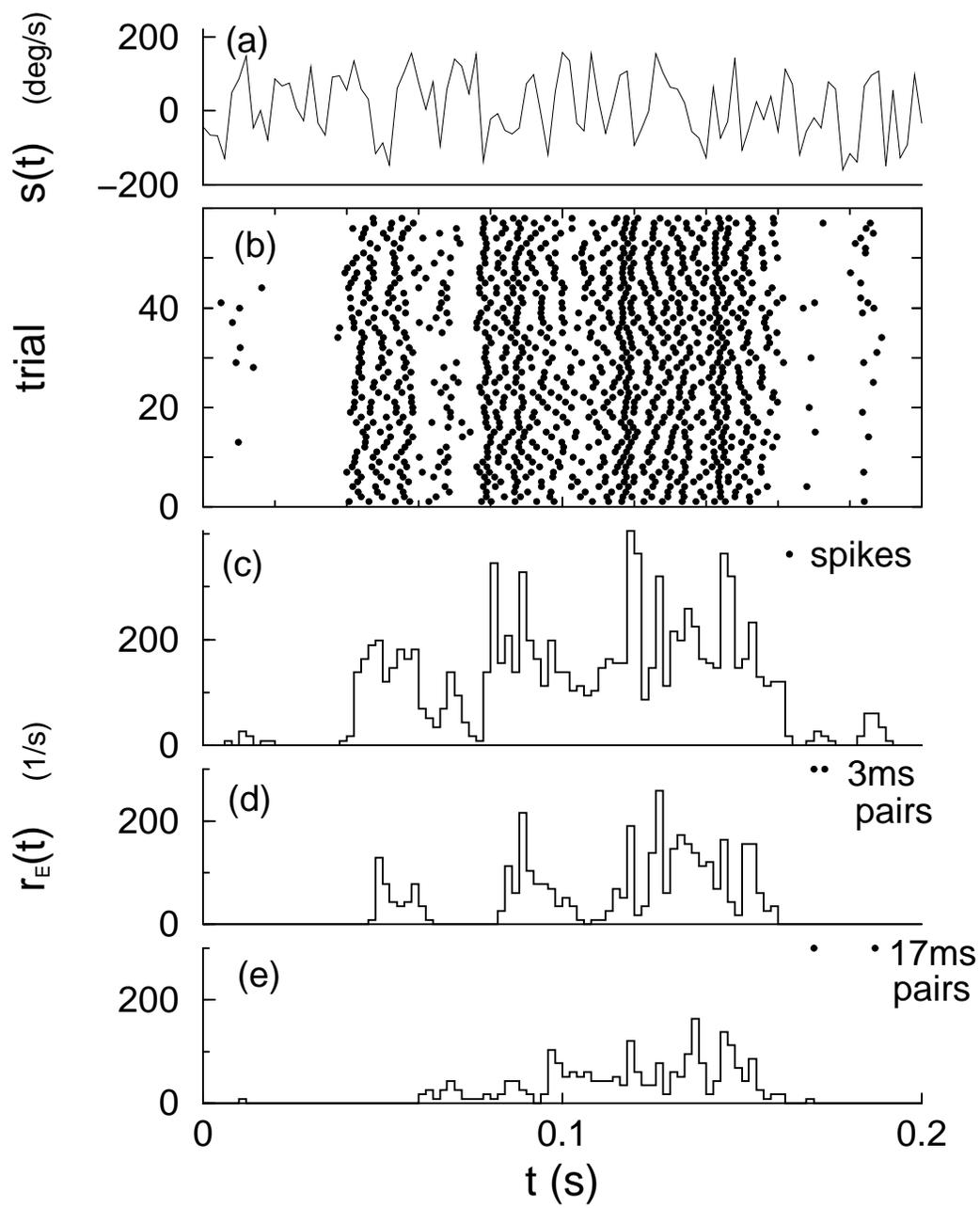
\noindent {\bf Fig. 1.}
Generalized event rates in the stimulus--conditional
response ensemble. A time dependent visual stimulus is shown to 
the fly (a), with the time axis defined to be zero at the 
beginning of the stimulus. This stimulus runs for
10 s, and is repeatedly presented 360 times. 
The responses of the H1 neuron to 60
repetitions are shown as a raster (b), in which each dot 
represents a single spike. From these responses, 
time dependent event rates
$r_E(t)$ are estimated: the firing rate (post-stimulus time 
histogram) (c); the  rate for spike pairs with interspike time 
$\tau = 3\pm 1$ ms (d) and for pairs
with $\tau = 17\pm 1$ ms (e).
These rates allow us to compute directly the
information transmitted by the events, using Eq. 
(\ref{final1eventinfo}).\\ \\

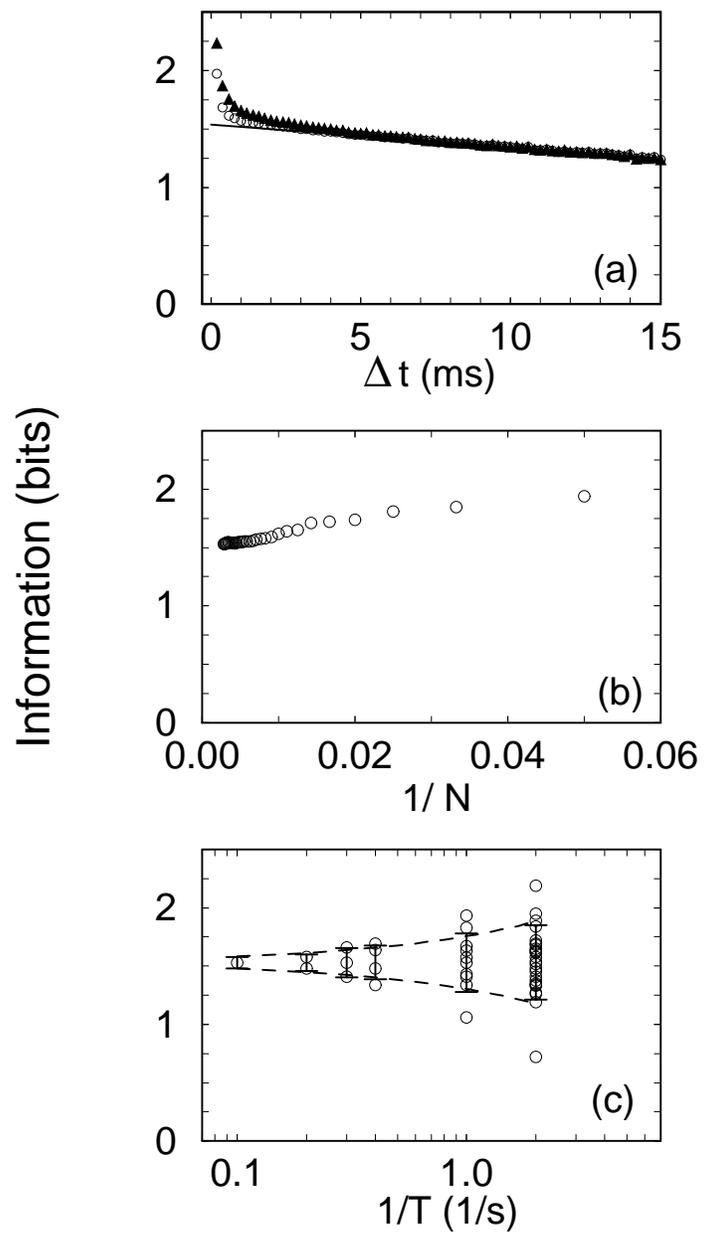
\noindent {\bf Fig. 2.}
\noindent Finite  size effects in the estimation of the
information conveyed by single spikes. (a) Information as a function
of the
bin size $\Delta t$ used for computing the time 
dependent rate $r(t)$ from all 350 repetitions (circles), and from
100 of the repetitions (filled triangles). 
A linear extrapolation to
the limit $\Delta t \to 0$ is shown for the case where all repetitions
were used (solid line).  
(b) Information as 
a function of the inverse number of repetitions $N$, for a fixed
bin size $\Delta t =2$ms. (c) Statistical error due to the finiteness
of the time segment of length $T$. Information shown as a function of the 
inverse time segment $1/T$, was obtained by dividing the full
10-sec segments into smaller segments of length $T$ (circles). 
For each such division, the
errorbars represent the standard deviation of the values obtained from
the different time intervals. These error bars should follow a
square-root law ($\sigma \propto 1/\sqrt{T}$) if the small segments are
independent. The dashed line shows the best power law fit to the
sequence of standard deviations, which extrapolates to an errorbar of
$\sigma\approx 0.05$ for the full 10-sec segment.\\ \\

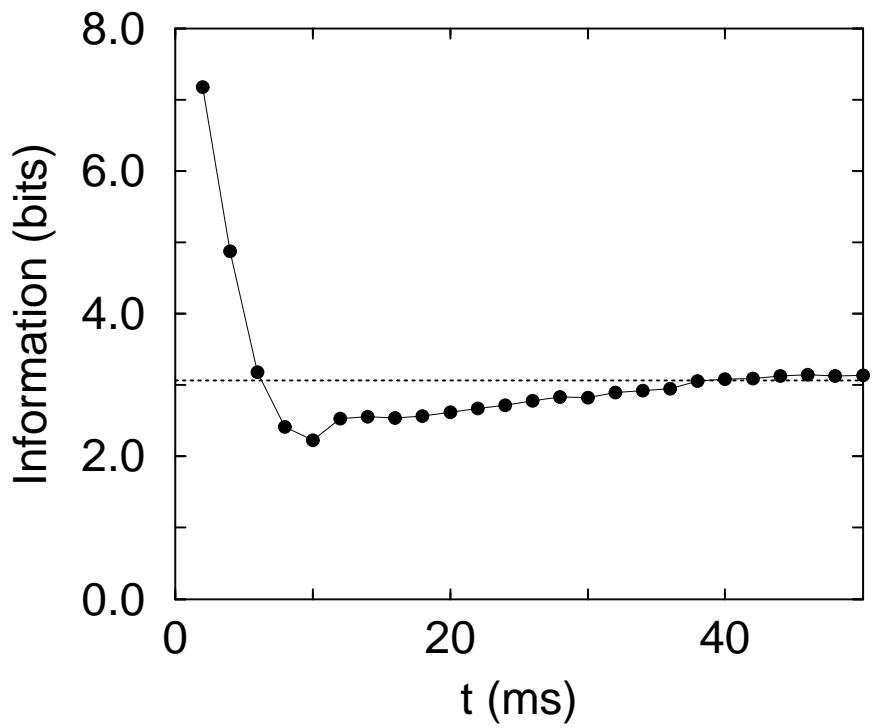
\noindent {\bf Fig. 3.}
Information about the signal transmitted by pairs of
spikes, computed from Eq. (\ref{final1eventinfo}), as a function of the time
separation between the two spikes.
The dotted line shows the information that would be transmitted
by the two spikes independently (twice the  single spike information).\\
\\

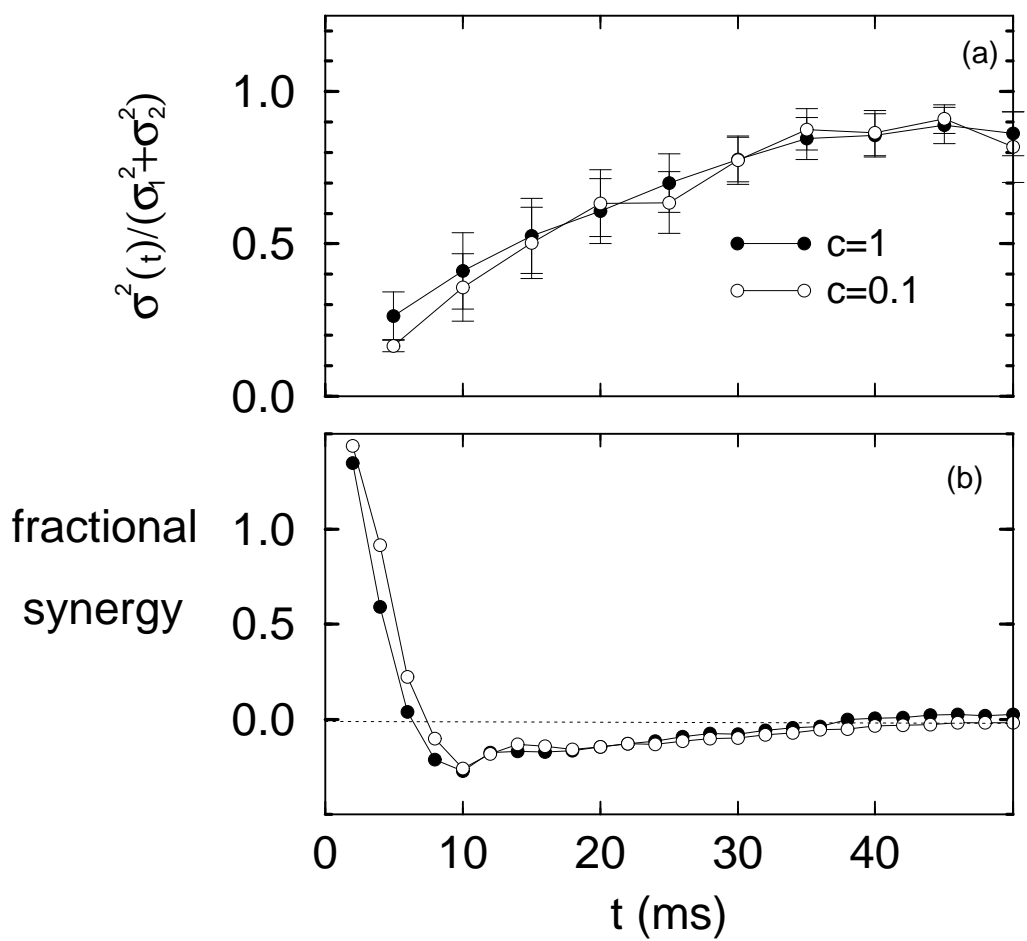
\noindent {\bf Fig. 4.}
(a) Ratio between the variance of interspike time and the sum of
variances
of the two spike times. Variances are measured across repeated
presentations of
same stimulus, as explained in the text. 
This ratio is plotted as a function of the
interspike time
$\tau$, for two experiments with different image contrast.
(b) Extra information conveyed cooperatively  by pairs of spikes,
expressed as a fraction of the information conveyed by the two spikes
independently. While the single spike information varies with contrast,
(1.5 bits/spike for c=0.1 compared to 1.3 bits/spike for c=1), the
fractional synergy is almost contrast independent.\\ \\

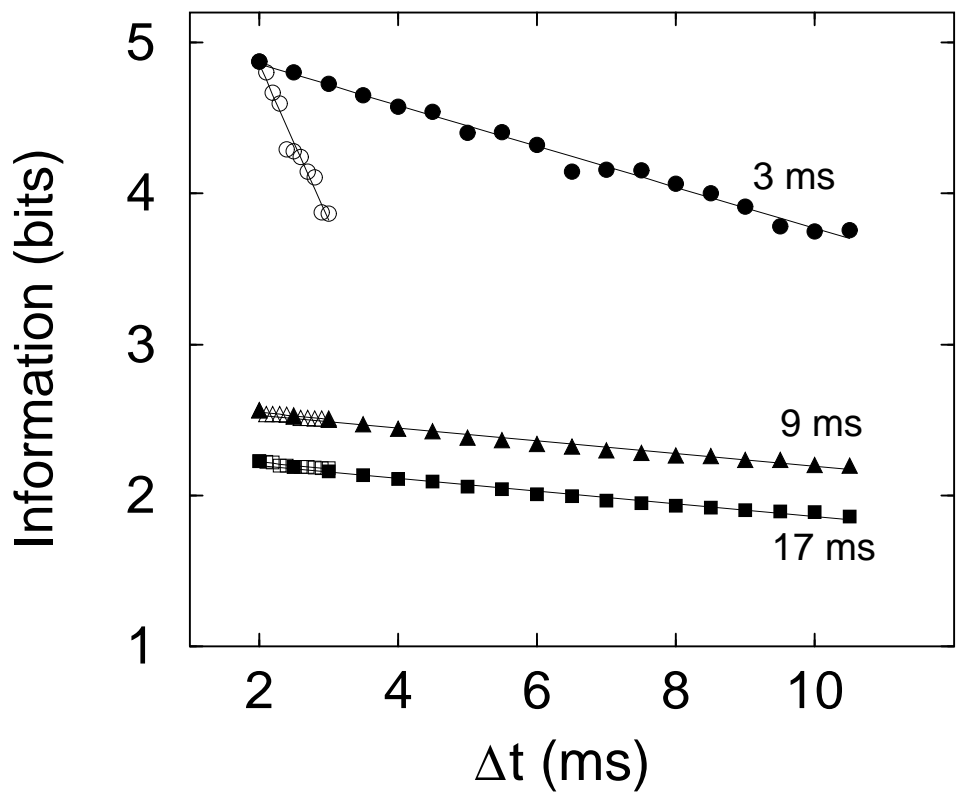
\noindent {\bf Fig. 5.}
Information conveyed by spike pairs  as a function of time resolution.
An event --pair of spikes-- can be described by two times:
the separation between 
spikes (relative time), and the occurrence time of the event
with respect to the 
stimulus (absolute time).  The information carried by the pair depends
on the time resolution in these two dimensions, both specified by the
bin size $\Delta t$. Open symbols are measurements of the information for
a fixed absolute-time resolution of 2 ms, and a variable relative-time
resolution $\Delta t$. Closed symbols correspond to a fixed
relative-time resolution of 2 ms, and a variable absolute-time
resolution $\Delta t$.
For short intervals, the sensitivity to coarsening of the relative time
resolution
is much greater than to coarsening of the absolute time
resolution 
In contrast, sensitivity to relative and absolute time
resolution is the same for the 
longer, nonsynergistic, interspike
separations.

\end{document}